# Metallicity effect in multi-dimensional SNIa nucleosynthesis

C. Travaglio[1,2], W. Hillebrandt[2] and M. Reinecke[2]

[1] Istituto Nazionale di Astrofisica (INAF) - Osservatorio Astronomico di Torino, Via Osservatorio 20, 10025 Pino Torinese (Torino), Italy
e-mail: travaglio@to.astro.it

[2] Max-Planck Institut für Astrophysik, Karl-Schwarzschild Strasse 1, D-85741 Garching bei München, Germany

**Abstract.** We investigate the metallicity effect (measured by the original $^{22}$Ne content) on the the detailed nucleosynthetic yields for 3D hydrodynamical simulations of the thermonuclear burning phase in type Ia supernovae (SNe Ia). Calculations are based on post-processes of the ejecta, using passively advected tracer particles, as explained in details by Travaglio et al. (2004). The nuclear reaction network employed in computing the explosive nucleosynthesis contains 383 nuclear species, ranging from neutrons, protons, and $\alpha$-particles to $^{98}$Mo. For this work we use the high resolution multi-point ignition (bubbles) model *b30_3d_768* (Travaglio et al. 2004 for the solar metallicity case), and we cover a metallicity range between $0.1\times Z_{\odot}$ up to $3\times Z_{\odot}$. We find a linear dependence of the $^{56}$Ni mass ejected on the progenitor's metallicity, with a variation in the $^{56}$Ni mass of $\sim 25\%$ in the metallicity range explored. Moreover, the largest variation in $^{56}$Ni occurs at metallicity greater than solar. Almost no variations are shown in the unburned material $^{12}$C and $^{16}$O. The largest metallicity effect is seen in the $\alpha$-elements. Finally, implications for the observed scatter in the peak luminosities of SNe Ia are also discussed.

## 1. Introduction

The understanding of the influence of an exploding white dwarf's initial composition on the nucleosynthesis, light curves, and spectra of Type Ia supernovae is an important tool to evaluate the origin of their observed diversity. It is widely accepted that SNe Ia are thermonuclear explosions of C+O white dwarfs, although the nature of the progenitor binary system and the details of the explosion mechanism are still under debate. Over the last decades, one-dimensional spherically symmetric models have been used to predict spectra, light curves and nucleosynthesis. Moreover, the dependence of the $^{56}$Ni ejected on the progenitor's metallicity as well as on the initial C/O composition has been investigated in literature (Höflich, Wheeler, & Thielemann 1998; Iwamoto et al. 1999; Umeda et al. 2000; Höflich et al. 2000; Dominguez, Höflich, & Straniero 2001; Timmes, Brown, & Truran 2003). More recently it has become possible to perform multidimensional simulations of exploding white dwarfs (see Reinecke, Hillebrandt, & Niemeyer 2002 and references therein; Gamezo et al. 2003). Also a detailed study of the nucleosynthesis using multi-dimensional SNIa models has been recently performed for solar metallicity initial composition (Travaglio et al. 2004). As described by Travaglio et al. 2004, the nucleosynthetic yields of multi-dimensional Eulerian hydrodynamic calculations of SNIa explosions have been obtained by post-processing the ejecta, using the density and temperature history of passively advected tracer particles.

Starting with the highest resolution pure deflagration SNIa model presented by Travaglio et al. (2004), *b30_3d_768* (a 3D model with ignition in 30 bubbles and grid size of $768^3$), we explore in this work the metallicity effect on the nucleosynthesis. In addition to the current metallicity study, we are also performing a detailed parameter study of the variation of the central density and of the initial carbon/oxygen ratio of the SNIa progenitor. This will be presented elsewhere (Röpke et al., in preparation). In Sect. 2 of this work we demonstrate that the mass of $^{56}$Ni depends linearly on the initial metallicity of the progenitor, in agreement with recent results from Timmes et al. (2003). We also discuss our results for the detailed nucleosynthetic composition of the SNIa models analyzed, and we compare them with the W7 calculations by Brachwitz et al. (2000) and Thielemann et al. (2003). Finally, in Sect. 3 we summarize our results.

## 2. Effects of variations in $^{22}$Ne progenitor's abundance

In order to typically simulate a solar metallicity SNIa, the initial white dwarf composition we use (mass fraction) 0.475 $M_{\odot}$ of $^{12}$C, 0.5 $M_{\odot}$ of $^{16}$O, and 0.025 $M_{\odot}$ of $^{22}$Ne. This is in agreement with the standard W7 initial composition (Iwamoto et al. 1999). As soon as the flame passes through the fuel, $^{12}$C,

*Send offprint requests to*: C. Travaglio

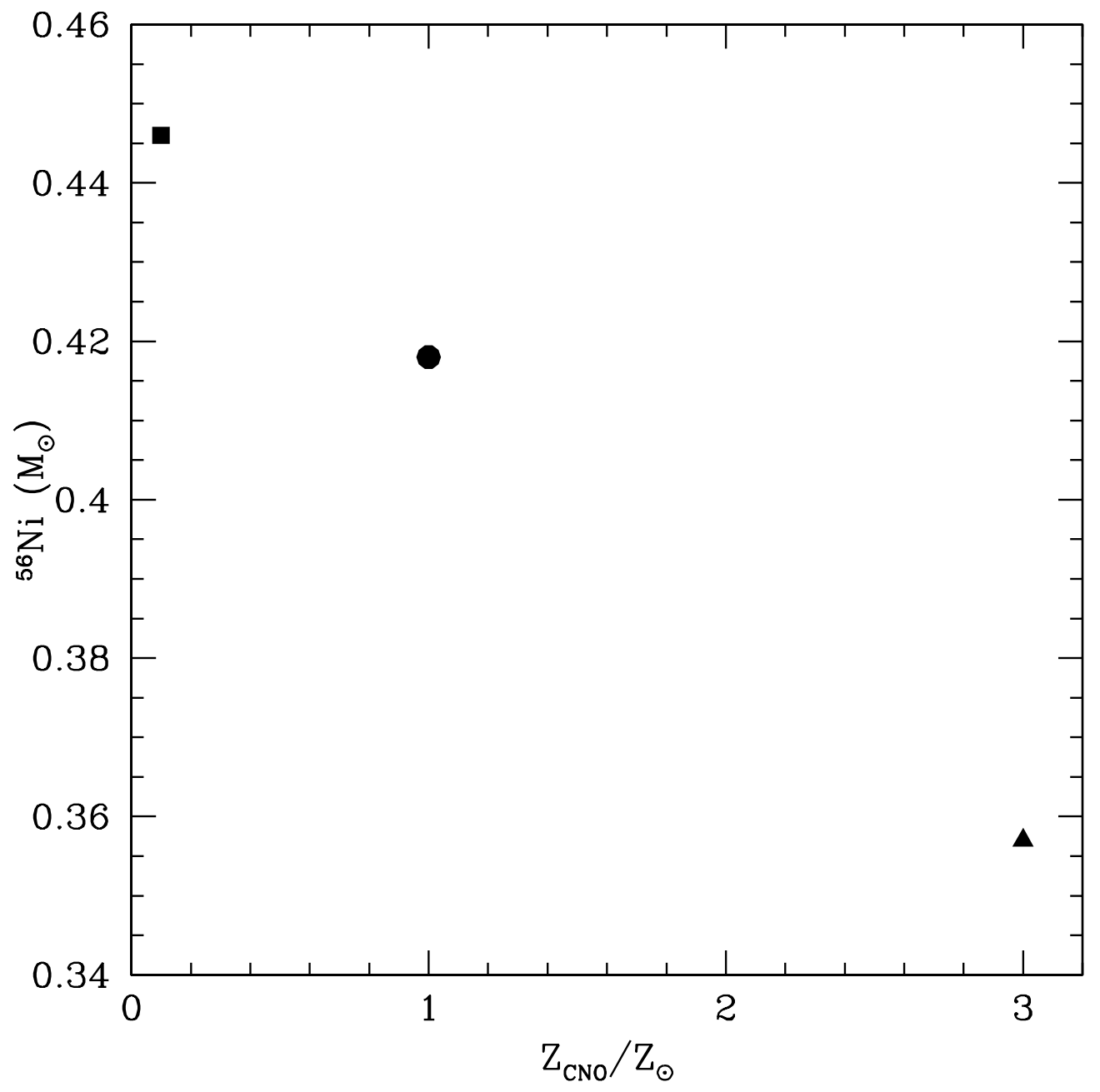

**Fig. 1.** $^{56}$Ni ejected mass by the *b30_3d_768* model as a function of the initial metallicity $Z_{CNO}$.

$^{16}$O, and $^{22}$Ne are converted to ashes with different compositions depending on the initial temperature and density. We simulate a metallicity effect changing the initial $^{22}$Ne abundance (and the $^{12}$C initial mass as a consequence). The reason is that the metallicity mainly affects the initial CNO abundances in a star. They are converted during pre-explosive H burning to $^{14}$N and He burning to $^{14}N(\alpha,\gamma)^{18}F(\beta^{+})^{18}O(\alpha,\gamma)^{22}Ne$ to heavier nuclei. Therefore a change of $^{22}$Ne abundance simulate the metallicity effect. We note that temperature and density profiles were calculated using the energy released by burning matter of solar metallicity. This seems to be a fair approximation because the metallicity is unlikely to influence the velocity of the flame front (which is independent of the microphysics in case of strong turbulence), and also the energy production depends only weekly on the metallicity. In fact, a mixture of higher $^{22}$Ne and less $^{12}$C abundance should give less energy, due to the differences in binding energies. However, this effect is small as long as the $^{22}$Ne abundance is on the order of a few percent only.

For the models mentioned above, we use the following initial composition: 0.4975 $M_\odot$ of $^{12}$C, 0.5 $M_\odot$ of $^{16}$O, and 0.0025 $M_\odot$ of $^{22}$Ne for the model *b30_3d_768_d10*; and 0.425 $M_\odot$ of $^{12}$C, 0.5 $M_\odot$ of $^{16}$O, and 0.075 $M_\odot$ of $^{22}$Ne for the model *b30_3d_768_p3*.

A complete view of our nucleosynthesis calculations for the models *b30_3d_768* (Travaglio et al. 2004), *b30_3d_768_d10* and *b30_3d_768_p3* is reported in Table 1 (synthesized masses for the main radioactive species from $^{22}$Na up to $^{63}$Ni), and in Table 2 (synthesized masses for all the stable isotopes up to $^{68}$Zn). In both Tables we also include (in Col. 2) for comparison the calculations for the W7 model (from Thielemann et al. 2003, and Brachwitz et al. 2000). In Fig. 2 we show the yields obtained for the models *b30_3d_768_d10* and for *b30_3d_768_p3*, normalized to the "standard" *b30_3d_768* case.

As a first interesting result, we find that the mass fraction of $^{56}$Ni depends linearly on the initial metallicity, that can be fitted by the simple following linear equation

$$M(^{56}Ni) \simeq 0.45 M_\odot - 0.031 Z_{CNO}/Z_\odot$$

Since $^{22}$Ne scales with $Z_{CNO}$, we scaled $Z_{CNO}$ rather than $Z_{Fe}$, and we use this notation for the whole paper. As one can see from Table 1, varying the metallicity from $Z_{CNO}$ = 0.1 to 3 $Z_\odot$ the $^{56}$Ni mass ejected changes of $\sim$ 25%. This is in a good agreement with the results presented by Timmes et al. (2003) (they found a variation of $^{56}$Ni mass ejected of $\sim$ 25% in a metallicity range 1/3 up to 3 $Z_\odot$). Also in agreement with Timmes et al. (1998) we find that the largest variation in the mass of $^{56}$Ni occurs at metallicity greater than solar ($\sim$15%). In contrast, previous investigations of the effect of variations of $^{22}$Ne by Höflich et al. (1998) and Iwamoto et al. (1999) do not agree on the $^{56}$Ni mass produced. Höflich et al. (1998) found that a metallicity variation from 0.1 to 10 $Z_\odot$ produces only a $\sim$4% variation in the $^{56}$Ni mass ejected. Instead Iwamoto et al. (1999) found that a variation of metallicity from zero to solar decreases the $^{56}$Ni mass of $\sim$10%. Much smaller $^{56}$Ni variations with metallicity can be explained as a difference in temperature profiles, i.e. if temperatures are higher in a large part of the inner white dwarf the electron captures could hide the metallicity effect. Alternatively, the delayed detonation effect produces a lot of $^{56}$Ni in the outer layers and the outcome should depend on the $^{22}$Ne distribution.

Still concerning our results for the $^{56}$Ni mass, the amplitude of its variations cannot account for all the observed variation in peak luminosity of SNIa (Pinto & Eastman 2001). And the observed scatter in the peak brightnesses may be even larger when more distant SNe are included (Hamuy et al. 1996; Riess et al. 1998; and more recently including Type Ia Supernova discoveries at $z > 1$ from the Hubble Space Telescope and from the Canada-France-Hawaii Telescope, Riess et al. 2004 and Barris et al. 2004, respectively).

Strong metallicity effects are also shown in the variations of the alpha elements (see Table 2), like Mg and Al isotopes. Instead almost no variations are shown in the Ti and V region. Moreover, interesting is the behavior of $^{54}$Fe, as also discussed in details by Höflich et al. (1998) and Timmes et al. (2003). As described more in details by Travaglio et al. (2004), when $T$ and $\rho$ are high enough, neutron-rich nuclei are built up due to electron captures, and $^{56}$Fe is partly replaced by $^{54}$Fe and $^{58}$Ni. We find that $^{56}$Fe is anti-correlated with $^{54}$Fe and $^{58}$Ni ejected, i.e. $^{54}$Fe and $^{58}$Ni increase with increasing metallicity. The largest effect is shown by the variation of $^{54}$Fe, a decrease of a factor of $\sim$3 for the model *b30_3d_768_d10*, and an increase of a factor of $\sim$2 for the model *b30_3d_768_p3*. Therefore subluminous SNe Ia will tend to have larger $^{54}$Fe/$^{56}$Fe ratios than brighter ones.

## 3. Summary and conclusions

In this letter we discussed the results of detailed nucleosynthesis calculations obtained from coupling 3D hydrodynam-ics of SN Ia explosion to post-processes of the ejecta using a tracer particle method. Nucleosynthesis and hydrodynamic calculations for the high resolution multi-point ignition model *b30_3d_768* discussed here, are explained in details by Travaglio et al. (2004). The purpose of the present work is to investigate the metallicity effects on the nucleosynthesis, obtained by changing the original $^{22}$Ne content. We presented here our results for two metallicities ($0.1Z_\odot$ and $3Z_\odot$, model *b30_3d_768_d10* and *b30_3d_768_p3* respectively), compared to the solar metallicity case *b30_3d_768*. We also discuss them in comparison with the standard W7 SNIa model (Iwamoto et al. 1999, Brachwitz et al. 2000, Thielemann et al. 2003). The approach is not fully consistent because the metallicity may also affect the C/O of the WD at explosion. However it has been shown recently (Roepke & Hillebrandt 2004) that the C/O ratio does not change the Ni-production and, thus, the peak luminosity of a type Ia supernovae by much if all other properties are kept constant. So we believe that our models catches the essential effect.

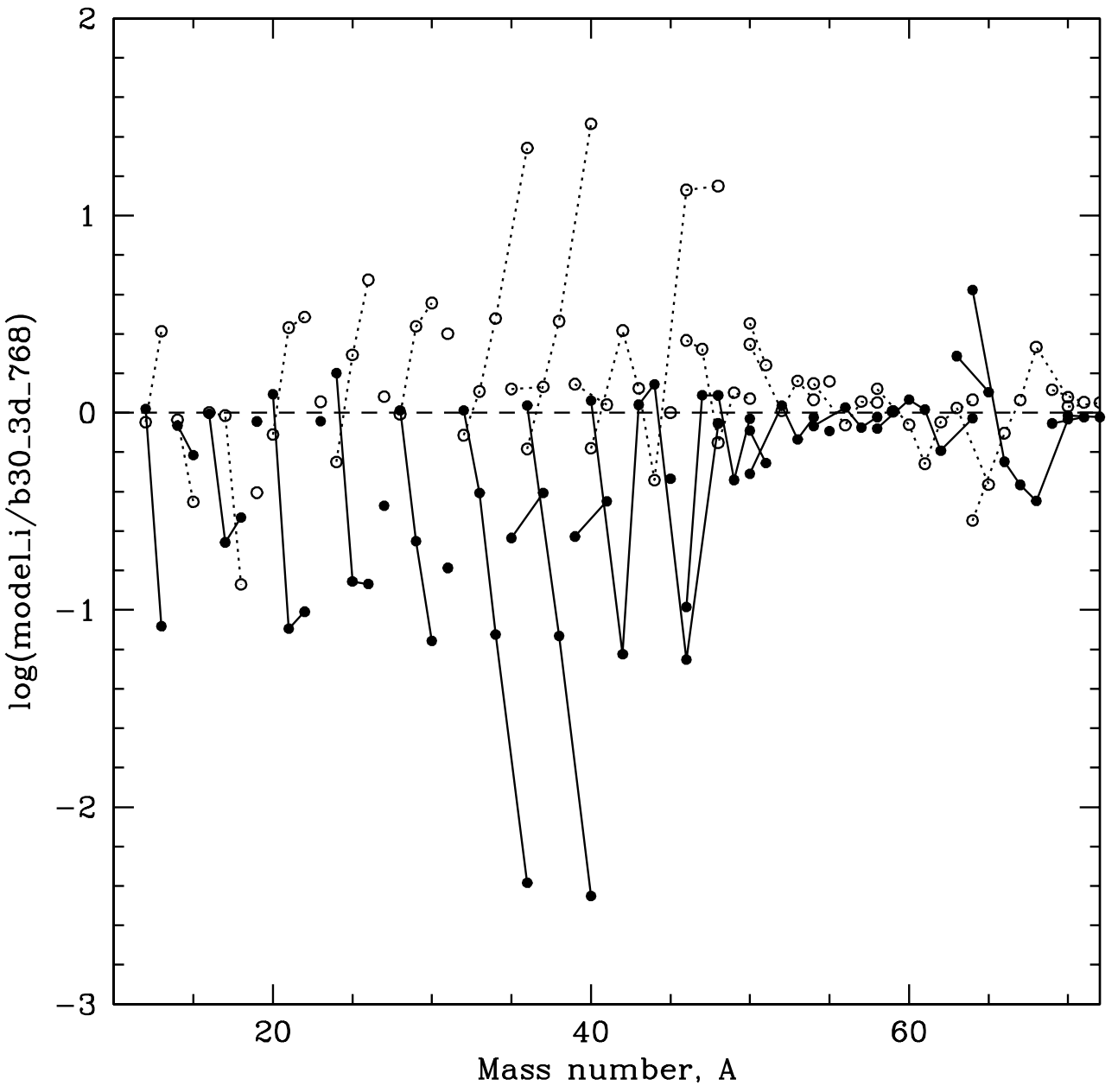

**Fig. 2.** Nucleosynthetic yields (in mass fraction) for the model *b30_3d_768* with $Z_{CNO} = 0.1Z_\odot$ (*solid lines and filled dots*) and for the model *b30_3d_768* with $Z_{CNO} = 3xZ_\odot$ (*dotted lines and open dots*). Both are normalized to the *b30_3d_768* model with solar metallicity.

We find that the $^{56}$Ni mass produced decreases linearly with metallicity, with a variation of $^{56}$Ni of ∼25% in the metallicity range explored. The largest variation (∼15%) is seen at $Z_{CNO} = 3Z_\odot$, the highest metallicity investigated in this study. We also discussed the behavior of all the other isotopes with a code that includes 383 isotopes. Interesting changes are shown for isotopes in the Mg and Al isotopes, and particular attention was dedicated to the $^{54}$Fe and $^{58}$Ni isotopes in comparison to $^{56}$Fe.

Metallicity effects on the nucleosynthesis using different hydrodynamical SN Ia models are currently in preparation (Röpke et al., in preparation). Also a detailed parameter study of the central density and of the initial carbon/oxygen ratio of the SN Ia progenitor with the effect on the nucleosynthesis will be presented by Röpke et al.

TABLE 1
SYNTHESIZED MASS ($M_\odot$) FOR RADIOACTIVE SPECIES IN SNIA MODELS

| Species | W7[a] | *b30_3d_768*[b] | *b30_3d_768_d10*[c] | *b30_3d_768_p3*[d] |
|---|---|---|---|---|
| $^{22}$Na | 1.73E-08 | 1.00E-07 | 1.69E-07 | 5.33E-08 |
| $^{26}$Al | 4.93E-07 | 4.47E-06 | 1.26E-06 | 3.61E-06 |
| $^{36}$Cl | 2.58E-06 | 6.32E-07 | 7.62E-07 | 1.16E-06 |
| $^{39}$Ar | 1.20E-08 | 2.24E-09 | 3.82E-08 | 1.11E-08 |
| $^{40}$K | 8.44E-08 | 1.23E-08 | 8.24E-10 | 1.75E-08 |
| $^{41}$Ca | 6.09E-06 | 2.40E-06 | 8.56E-07 | 2.62E-06 |
| $^{44}$Ti | 7.94E-06 | 3.61E-06 | 5.04E-06 | 1.59E-06 |
| $^{48}$V | 4.95E-08 | 7.68E-05 | 9.42E-05 | 5.38E-05 |
| $^{49}$V | 1.52E-07 | 5.78E-06 | 2.64E-06 | 7.31E-06 |
| $^{53}$Mn | 2.77E-04 | 4.79E-04 | 3.50E-04 | 6.96E-04 |
| $^{60}$Fe | 7.52E-07 | 4.52E-10 | 4.20E-10 | 5.37E-10 |
| $^{56}$Co | 1.44E-04 | 1.32E-04 | 1.20E-04 | 1.65E-04 |
| $^{57}$Co | 1.48E-03 | 1.15E-03 | 1.11E-03 | 1.27E-03 |
| $^{60}$Co | 4.22E-07 | 2.66E-08 | 2.54E-08 | 2.97E-08 |
| $^{56}$Ni | 5.86E-01 | 4.18E-01 | 4.46E-01 | 3.57E-01 |
| $^{57}$Ni | 2.27E-02 | 1.74E-02 | 1.45E-02 | 1.99E-02 |
| $^{59}$Ni | 6.71E-04 | 7.11E-04 | 7.27E-04 | 7.22E-04 |
| $^{63}$Ni | 8.00E-07 | 2.22E-08 | 2.10E-08 | 2.52E-08 |

[a] Brachwitz et al. (2001)
[b] Travaglio et al. (2004)
[c] This work, with $Z_{CNO} = 0.1 \times Z_\odot$
[d] This work, with $Z_{CNO} = 3 \times Z_\odot$

*Acknowledgements.* C.T. thanks the Alexander von Humboldt Foundation, the Federal Ministry of Education and Research, and the Programme for Investment in the Future (ZIP) of the German Government, and the Max-Planck Institute für Astrophysik (Garching bei München), for their financial support. This work was supported in part by the the "Sonderforschungsbereich 375-95 für Astro-Teilchenphysik" der Deutschen Forschungsgemeinschaft and the European Commission under grant HPRN-CT-2002-00303.

TABLE 2
SYNTHESIZED MASS ($M_\odot$) IN SNIA MODELS

| Species | W7[a] | *b30_3d_768*[b] | *b30_3d_768_d10*[c] | *b30_3d_768_p3*[d] |
|---|---|---|---|---|
| $^{12}$C | 5.04E-02 | 2.78E-01 | 2.91E-01 | 2.49E-01 |
| $^{13}$C | 1.07E-06 | 3.98E-06 | 3.29E-07 | 1.03E-05 |
| $^{14}$N | 4.94E-07 | 2.76E-04 | 2.37E-04 | 2.54E-04 |
| $^{15}$N | 1.25E-09 | 1.23E-06 | 7.53E-07 | 4.36E-07 |
| $^{16}$O | 1.40E-01 | 3.39E-01 | 3.35E-01 | 3.40E-01 |
| $^{17}$O | 3.05E-08 | 1.31E-06 | 2.88E-07 | 1.27E-06 |
| $^{18}$O | 7.25E-10 | 1.01E-05 | 2.99E-06 | 1.37E-06 |
| $^{19}$F | 5.72E-10 | 2.84E-08 | 2.56E-08 | 1.12E-08 |
| $^{20}$Ne | 1.97E-03 | 6.28E-03 | 7.79E-03 | 4.87E-03 |
| $^{21}$Ne | 8.51E-06 | 2.16E-05 | 1.73E-06 | 5.83E-05 |
| $^{22}$Ne | 2.27E-03 | 1.42E-02 | 1.39E-03 | 4.34E-02 |
| $^{23}$Na | 6.20E-05 | 8.65E-04 | 7.84E-04 | 9.83E-04 |
| $^{24}$Mg | 1.31E-02 | 7.53E-03 | 1.19E-02 | 4.24E-03 |
| $^{25}$Mg | 4.71E-05 | 5.13E-04 | 7.16E-05 | 1.01E-03 |
| $^{26}$Mg | 3.31E-05 | 1.81E-04 | 2.45E-05 | 8.54E-04 |
| $^{27}$Al | 8.17E-04 | 5.85E-04 | 1.97E-04 | 7.06E-04 |
| $^{28}$Si | 1.52E-01 | 5.39E-02 | 5.55E-02 | 5.31E-02 |
| $^{29}$Si | 7.97E-04 | 5.61E-04 | 1.25E-04 | 1.54E-03 |
| $^{30}$Si | 1.43E-03 | 8.03E-04 | 5.60E-05 | 2.89E-03 |
| $^{31}$P | 3.15E-04 | 1.72E-04 | 2.80E-05 | 4.32E-04 |
| $^{32}$S | 8.45E-02 | 2.62E-02 | 2.70E-02 | 2.02E-02 |
| $^{33}$S | 4.11E-04 | 1.21E-04 | 4.76E-05 | 1.56E-04 |
| $^{34}$S | 1.72E-03 | 1.04E-03 | 7.86E-05 | 3.14E-03 |
| $^{36}$S | 2.86E-07 | 1.53E-07 | 6.33E-10 | 3.38E-06 |
| $^{35}$Cl | 1.26E-04 | 4.58E-05 | 1.06E-05 | 6.04E-05 |
| $^{37}$Cl | 3.61E-05 | 1.21E-05 | 4.74E-06 | 1.64E-05 |
| $^{36}$Ar | 1.49E-02 | 4.24E-03 | 4.62E-03 | 2.78E-03 |
| $^{38}$Ar | 8.37E-04 | 5.59E-04 | 4.12E-05 | 1.63E-03 |
| $^{40}$Ar | 1.38E-08 | 1.91E-09 | 4.76E-12 | 5.58E-08 |
| $^{39}$K | 6.81E-05 | 3.24E-05 | 7.62E-06 | 4.52E-05 |
| $^{41}$K | 6.03E-06 | 2.41E-06 | 8.56E-07 | 2.64E-06 |
| $^{40}$Ca | 1.21E-02 | 3.59E-03 | 4.14E-03 | 2.37E-03 |
| $^{42}$Ca | 2.48E-05 | 1.58E-05 | 9.44E-07 | 4.13E-05 |
| $^{43}$Ca | 1.07E-07 | 5.10E-08 | 5.60E-08 | 6.78E-08 |
| $^{44}$Ca | 9.62E-06 | 3.61E-06 | 5.04E-06 | 1.65E-06 |
| $^{46}$Ca | 2.44E-09 | 8.53E-12 | 4.78E-13 | 1.15E-10 |
| $^{48}$Ca | 1.21E-12 | 4.01E-15 | 3.54E-15 | 5.67E-14 |
| $^{45}$Sc | 2.17E-07 | 6.47E-08 | 2.99E-08 | 6.48E-08 |
| $^{46}$Ti | 1.16E-05 | 6.62E-06 | 6.85E-07 | 1.54E-05 |
| $^{47}$Ti | 5.45E-07 | 2.64E-07 | 3.24E-07 | 5.56E-07 |
| $^{48}$Ti | 2.07E-04 | 7.69E-05 | 9.42E-05 | 5.43E-05 |
| $^{49}$Ti | 1.59E-05 | 5.78E-06 | 2.64E-06 | 7.32E-06 |
| $^{50}$Ti | 1.62E-06 | 2.67E-07 | 2.49E-07 | 3.15E-07 |
| $^{50}$V | 4.58E-09 | 2.66E-09 | 2.16E-09 | 7.56E-09 |
| $^{51}$V | 3.95E-05 | 1.95E-05 | 1.08E-05 | 3.40E-05 |
| $^{50}$Cr | 2.23E-04 | 1.19E-04 | 5.85E-05 | 2.65E-04 |
| $^{52}$Cr | 4.52E-03 | 2.58E-03 | 2.80E-03 | 2.64E-03 |
| $^{53}$Cr | 6.49E-04 | 4.83E-04 | 3.54E-04 | 7.01E-04 |
| $^{54}$Cr | 3.04E-05 | 1.22E-05 | 1.16E-05 | 1.42E-05 |
| $^{55}$Mn | 6.54E-03 | 6.38E-03 | 5.15E-03 | 9.19E-03 |
| $^{54}$Fe | 7.49E-02 | 7.33E-02 | 2.27E-02 | 1.03E-01 |
| $^{56}$Fe | 6.69E-01 | 4.39E-01 | 4.67E-01 | 3.80E-01 |
| $^{57}$Fe | 2.52E-02 | 1.86E-02 | 1.56E-02 | 2.12E-02 |
| $^{58}$Fe | 1.74E-04 | 1.05E-04 | 9.96E-05 | 1.18E-04 |
| $^{59}$Co | 7.66E-04 | 7.33E-04 | 7.48E-04 | 7.46E-04 |
| $^{58}$Ni | 1.02E-01 | 9.66E-02 | 8.03E-02 | 1.28E-01 |
| $^{60}$Ni | 9.22E-03 | 7.73E-03 | 9.02E-03 | 6.73E-03 |
| $^{61}$Ni | 2.69E-04 | 1.13E-04 | 1.17E-04 | 6.22E-05 |
| $^{62}$Ni | 2.31E-03 | 1.12E-03 | 7.18E-04 | 1.00E-03 |
| $^{64}$Ni | 1.84E-07 | 5.29E-08 | 4.96E-08 | 6.15E-08 |
| $^{63}$Cu | 1.59E-06 | 9.56E-07 | 1.85E-06 | 1.01E-06 |
| $^{65}$Cu | 7.72E-07 | 3.77E-07 | 4.80E-07 | 1.63E-07 |
| $^{64}$Zn | 1.50E-05 | 6.78E-06 | 2.85E-05 | 1.92E-06 |
| $^{66}$Zn | 1.31E-08 | 1.16E-05 | 6.57E-06 | 9.16E-06 |
| $^{67}$Zn | 1.18E-11 | 7.96E-09 | 3.43E-09 | 9.22E-09 |
| $^{68}$Zn | 2.66E-10 | 5.26E-09 | 1.88E-09 | 1.13E-08 |

[a] Thielemann et al. (2003)
[b] Travaglio et al. (2004)
[c] This work, with $Z_{CNO}$ =0.1×$Z_\odot$
[d] This work, with $Z_{CNO}$ =3×$Z_\odot$